\input harvmac
\input epsf

\def\Title#1#2{\rightline{#1}\ifx\answ\bigans\nopagenumbers\pageno0\vskip1in
\else\pageno1\vskip.8in\fi \centerline{\titlefont #2}\vskip .5in}

\font\ticp=cmcsc10

\def\pp{\partial}
\def\({\left (}
\def\){\right )}
\def\[{\left [}
\def\]{\right ]}

\def\h{\hat}
\gdef\journal#1, #2, #3, 19#4#5{{\sl #1~}{\bf #2}, #3 (19#4#5)}

\lref\juan{J. Maldacena, hep-th/9607235.}
\lref\gary{F. Dowker, J. Gauntlett, G. Gibbons, and G. Horowitz, \journal Phys. Rev., D53, 7115, 1996, hep-th/9512154.}
\lref\joe{J. Polchinski, \journal Phys. Rev. Lett., 75, 4724, 1995, hep-th/9510017.}
\lref\lifsh{G. Lifschytz, \journal Phys. Lett., B388, 720, 1996, hep-th/9604156.}
\lref\gibbons{G. Gibbons, \journal Commun. math. Phys., 44, 245, 1975.}
\lref\garyandjoe{G. Horowitz and J. Polchinski, hep-th/9612146.}
\lref\tasi{J. Polchinski, hep-th/9611050.}
\lref\douglas{M. Douglas, D. Kabat, P. Pouliot, and S. Shenker, \journal Nucl. Phys., B485, 85, 1997, hep-th/9608024.}
\lref\joebook{J. Polchinski, Joe's Big Book of String, Unpublished}
\lref\introa{A. Strominger and C. Vafa, \journal Phys. Lett., B379, 99, 1996, hep-th/9601029.}
\lref\introb{J. Breckenridge, R. Myers, A. Peet and C. Vafa, \journal Phys. Lett., B391, 93, 1997, hep-th/9602065.}
\lref\introc{J. Maldacena and A. Strominger, \journal Phys. Rev. Lett., 77, 428, 1996, hep-th/9603060; C. Johnson, R.Khuri, and R. Myers, \journal Phys. Lett., B378, 78, 1996, hep-th/9603061.}
\lref\introd{C. Callan and J. Maldacena, \journal Nucl. Phys., B472, 591, 1996, hep-th/9602043.}
\lref\introe{G. Horowitz and A. Strominger, \journal Phys. Rev. Lett., 77, 2368, 1996, hep-th/9602051.}
\lref\introf{J. Breckenridge, D. Lowe, R. Myers, A. Peet, A. Strominger and C. Vafa, hep-th/9603078.}
\lref\introg{G. Horowitz, D. Lowe, and J. Maldacena, \journal Phys. Rev. Lett., 77, 430, 1996, hep-th/9603195.}
\lref\introh{A. Dabholkar, hep-th/9702050.}
\lref\introi{G. Horowitz, J. Maldacena, and A. Strominger, hep-th/9603109.}
\lref\gibwilt{G. Gibbons and D. Wiltshire, \journal Ann. Phys., 167, 201, 1986.}
\lref\ortin{R. R. Khuri and T. Ortin, \journal Phys. Lett., B373, 56, 1996, hep-th/9512178.}

\Title{\vbox{\baselineskip12pt\hbox{hep-th/9705054}}}
{\vbox{
{\centerline {Statistical Entropy of an Extremal Black Hole}}
{\centerline {with 0- and 6-Brane Charge}}
}}
\centerline{\ticp Harrison J. Sheinblatt}
\vskip.1in
\centerline{\sl Department of Physics, University of California,
Santa Barbara, CA 93106}
\centerline{\it hjs@cosmic.physics.ucsb.edu}

\bigskip
\centerline{\bf Abstract}

A black hole solution to low energy type IIA string theory which is
extremal, non-supersymmetric, and carries 0- and 6-brane charge 
is presented.  For large values of the charges it is metastable and a
corresponding D-brane picture can be found.  The mass and statistical
entropy of the two descriptions agree at a correspondence point up to
factors of order one, providing more evidence that the correspondence 
principle for black 
holes and strings of Horowitz and Polchinski may be extended to 
include black holes with more than one Ramond-Ramond charge.

\Date{}

\baselineskip=16pt

\newsec{Introduction}

Over the past year, the entropies of several varieties of black holes in
string theory have
been explained on a microscopic level in terms of configurations of
D-branes and strings.  When the string coupling, $g$, is taken to zero, a
black hole solution to a low-energy effective string theory becomes a
weakly coupled system described by D-branes and strings
in flat space.  For BPS black holes, a non-renormalization theorem fixes the
degrees of freedom as the string coupling is varied, so that the statistical
entropy computed using the small $g$ picture exactly matches the 
Bekenstein-Hawking entropy of the black hole.  Several of these black holes
have been studied, and the small string coupling limit of all of them 
correctly describes the microscopic origin
of the entropy \refs{\introa, \introb, \introc}.

Even more surprising is that this microscopic description of black holes
has been shown to correctly describe the entropy of certain non-BPS black
holes.  In particular, some near-extremal black holes \refs{\introd, \introe,
\introf} have been analyzed, as well as some black holes which are far from
BPS-states \refs{\introg, \introh}.  It is unclear in
these cases precisely why this procedure is successful.

Recently a less exact but more widely applicable correspondence principle
between the black hole and string pictures has been formulated \garyandjoe . 
The correspondence principle states that the mass and entropy of the 
two pictures
should match only at a particular value of the string coupling: its value
when, for fixed charges, the curvature at the black hole horizon is of order
one in string units.  This correspondence point is where the low energy 
effective theory breaks down; the string size is of the order of 
the Schwarzschild radius.  Since the precise value of $g$ at which the
transition to strong coupling occurs is unknown up to factors of order one,
the mass and entropy of the two pictures must match only up to factors of
order one.  This was shown to apply to all black holes with less than two
Ramond-Ramond (RR) charges, and an example with two RR charges was shown to
work, suggesting that the correspondence principle may hold for all black
holes.

In this paper a black hole solution to the low energy effective theory of type
IIA string theory is presented.  It is extremal, non-supersymmetric,
and carries 0- and 6-brane charge.
Despite the fact that 0- and 6-branes have been shown to repell each other,
they do form such bound states.  Furthermore, the correspondence principle
will be shown to hold in this case, providing additional evidence that it is 
valid for black holes with two RR charges.

The paper is organized as follows.  In Section 2 units and conventions are 
explained, and the black hole solution
is presented in Section 3.  In Section 4 interactions between 0- and 6-branes
are explored.  A definition of entropy is provided in the next section,
followed by a D-brane description of the black hole in Section 6.  Some
final remarks are then discussed.

\newsec{Units, Conventions, and Quantization}

The bosonic part of type IIA string theory has the low energy effective action
\eqn\twoaact{S = {{1}\over{\(2 \pi \)^7 g^2}} \int {d^{10}x\sqrt{-G}\[e^{-2\phi} \(R + 4 \( \nabla \phi \)^2 \) - g^2 F^2 \]}} 
where all matter terms but the Ramond-Ramond (RR) 2-form field strength, $F$, 
have been set to zero.  The remaining terms are the string metric, $G$, and 
the dilaton, $\phi$, where $\phi\rightarrow 0$ at spatial infinity.  These 
quantities are expressed in string units: $\sqrt{\alpha '} = 1$.  $g$ is the 
string coupling, and the ten-dimensional Newton's constant, $G^{10}_{N}$, 
is set to $8 \pi^6 g^2$ in accordance with S-duality considerations \juan\ . 

A solution to the equations of motion for this action can be obtained from a 
solution of the equations of motion of the 11-D supergravity action
\eqn\elevact{ {}^{11}S = {{1}\over{\(2 \pi \)^8 g^3}} \int {d^{11}x \sqrt{-^{11}g} \  {}^{11}R}}
of the form
\eqn\edmf{ {}^{11}ds^2 = e^{4\phi / 3} \(dx^{11} + 2 g A_{\mu} dx^{\mu} \)^2 + e^{-2 \phi / 3} \  {}^{10}ds^2}
by compactification in the $x^{11}$ direction provided 
$\partial\over{\partial x^{11}}$ is a Killing vector of the 
solution \gary\ .  The dilaton, $\phi$, gauge potential, $A_{\mu}$, and the 
10-dimensional string metric, ${}^{10}ds^2$, can be read off from the form 
of \edmf .  The radius of the compactified $x^{11}$ direction is required
to be $g$, the 
string coupling, so that $\phi \rightarrow 0$ at spatial infinity.
The fields in the 11-dimensional solution are then in 
10-dimensional string units.

In the 10-dimensional solution, electric charge under the field strength $F$ 
is carried by 0-dimensional objects which have RR charge: 0-branes.  
Likewise, objects with magnetic charge under $F$ are 6-branes \joe\ .  
Thus these charges should be quantized.  One way to determine this 
quantization is to note how these charges arise from the 11-dimensional 
solution.  Electric charge comes from momentum in the $x^{11}$ direction 
of ${}^{11}ds^2$ which, for compact $x^{11}$ must be quantized in units of 
inverse radius, $1/g$.  Magnetic charge comes from a monopole or topological 
charge; discrete values are required to remove possible conical 
singularities in the 11-dimensional solution.

The 11-dimensional solution representing a 0-brane upon reduction to 
10-dimensions can be written in the form of \edmf\ with
\eqn\zerobrane{\eqalign{ {}^{10}ds^2  &= -\(1+{c\over{r^7}}\)^{-1/2} dt^2 + \(1+{c\over{r^7}}\)^{1/2}\[dr^2 + r^2d\Omega^8\] \cr A_{\mu}dx^{\mu} &= {1\over{2}}{c\over{g}}{1\over{r^7+c}}dt \cr e^{4 \phi/3} &= 1+{c\over{r^7}}. }}
Setting the momentum in the $x^{11}$ direction to $1/g$ requires the 
constant $c$ to be $c = 15\(2\pi\)^2 g$.  This implies that in 10-dimensions 
the integer normalized 0-brane charge is
\eqn\zbc{Q_{o} = {1\over{2^6\pi^7}}\int_{S^8, r\rightarrow \infty} *F}
where $*$ is the 10-D Hodge dual and the mass of one 0-brane is
\eqn\zbm{M_{0} = {1\over{g}}.}

To find the integer normalized magnetic charge, consider the 11-dimensional 
supergravity solution of the form \edmf\ with
\eqn\sixbrane{\eqalign{ {}^{10}ds^2 &= \(1+{m\over{r}}\)^{-1/2}\[-dt^2 + dy_{i}dy^{i}\] +\(1+{m\over{r}}\)^{1/2}\[dr^2+r^2d\Omega^2\] \cr A_{\mu}dx^{\mu} &= {m\over{2g}}\(1-\cos{\theta}\)d\h{\phi} \cr e^{4\phi/3} &= 1+{m\over{r}},}}
which, after reduction 10-dimensions, becomes the 6-brane solution.  Here 
$\h{\phi}$ is the azimuthal 
angle of the two-spheres and the $y_{i}, i=1...6$ are coordinates along 
the 6-brane.  Consider the $y_i$ directions compactified upon a 6-torus with 
volume $V^6$ at $r \rightarrow \infty$ for convenience.  Making this 
solution free of singularities requires a period for $x^{11}$ of 
$\Delta x^{11} = 4 \pi m$ which means $m = {1\over{2}}g$.  Thus the integer 
normalized 6-brane charge is found to be
\eqn\sbc{Q_{6} = {1\over{\pi}}\int_{T^6 \times S^2} F}
and the mass of a single 6-brane is
\eqn\sbm{M_{6} = {1\over{g}} {V^6\over{\(2\pi\)^6}}.}

\newsec{Extremal Black Hole Solution}

A solution to \elevact\ in the form of \edmf\ which after reduction to 10-D is
an extremal black hole with 0- and 6-brane charge is
\eqn\blackhole{\eqalign{ {}^{10}ds^2 &= -\(1 - {{qg}\over{r}}\)^2dt^2 + \(1 - {{qg}\over{r}}\)^{-2}dr^2 + r^2d\Omega^2 + dy_{i}dy^{i} \cr A_{\mu}dx^{\mu} &= {1\over{\sqrt{2}}}q\[ {1\over{r}}dt + \(1 - \cos{\theta}\)d\h{\phi}\] \cr e^{4\phi/3} &= 1.}}
$\h{\phi}$ is the azimuthal angle of the two-sphere and $y_{i}, i=1...6$ are 
flat directions which will be considered compactified on a six-torus of 
volume $V^6$.  Note that the dilaton, $\phi$, is constant and there is only 
one parameter, $q$.  This 11-D solution was previously found in \ortin\ where 
a different compactification to 10-D was considered.

The mass, integer normalized 0- and 6-brane charges, and the 
Bekenstein-Hawking entropy can be determined in terms of the parameter $q$ 
to be
\eqn\mandc{\eqalign{M &= {1\over{g}} {V^6\over{\(2\pi\)^6}} 8q \cr Q_{o} &= {q\over{\sqrt{2}}} {V^6\over{\(2\pi\)^6}} 4 \cr Q_{6} &= {q\over{\sqrt{2}}} 4 \cr S_{BH} &= 8 \pi q^2 {V^6\over{\(2\pi\)^6}}.}}
The mass of the black hole can be expressed in terms of the integer 
normalized charges, the mass of a single 0-brane \zbm, and the mass of a 
single 6-brane \sbm :
\eqn\bhmass{M = \sqrt{2} \(Q_{o} M_{o} + Q_{6} M_{6}\).}
In terms of the integer normalized charges, the Bekenstein-Hawking entropy is
\eqn\bhentropy{S_{BH} = \pi Q_{0} Q_{6}.}
$Q_{0}$ and $Q_{6}$ are related by $Q_{0}/Q_{6} = V^6/\(2\pi\)^6$.  One can
consider arriving at this solution by directly solving the 10-D field
equations.  Then, requiring that the dilaton be constant restricts the charges
to be related in this way, and choosing an extremal solution fixes the mass
in terms of the charges.  For
$g Q_0 >> 1$ the solution has small curvature everywhere outside
the horizon, so the low energy effective description is valid in the region
of interest.

Several more features of this solution are worthy of note.  First, the mass is 
greater than that of the constituent D-branes by a factor of $\sqrt{2}$.  
That the mass of a bound state of 0- and 6-branes is greater than the mass 
of separate 0- and 6-branes is not unusual given that 0- and 6-branes repell 
at large distances \lifsh ; energy must be added to force the D-branes 
together.  However, because the ``ground state'' of infinitely separated 
D-branes exists, it is expected that this black hole should be unstable to 
decay of 0- and 6-branes.

Even extreme black holes, which have zero Hawking temperature, can decay
semiclassically if there exist particles in the theory with greater charge 
than mass \gibbons .  The added gauge interaction allows negative energy 
orbits outside the event horizon producing an effective ergosphere.  
Because the black hole considered here has no dilaton charge, this is the only
requirement for there to be radiation.  In 10-D 
planck units, both the 0-brane and 6-brane have charge equal to twice their 
mass, so the black hole can radiate both kinds of D-branes.  For large black 
holes, where the black hole charges are effectively unchanged by the 
radiation, a WKB approximation \gibbons\ estimates the decay rate into 
either type of D-brane to be 
$e^{-k Q}$ where $Q$ is $Q_0$ $\( Q_6 \)$ for 0-brane (6-brane) radiation 
and $k$ is a constant of order one.
Thus, large black holes of this type are 
metastable.

\newsec{Effective Description of 0-6 String}

In order to construct a D-brane description of this black hole one needs to
understand the simpler system of one 0-brane and one 6-brane with small
separation.  It has been
argued that the short distance physics of such D-brane configurations is
best described by massless open string modes connecting the branes \douglas\ .
Thus, understanding this simpler system reduces to understanding the 0-6
open string stretched between the branes.

The effective potential between two D-branes can be found from the open
string loop amplitude in the D-brane background which has been calculated
for stationary D-branes in \lifsh .  The short distance limit of this
amplitude is the infinite world sheet time limit of the integral.  For the
0-6 brane system this yields an effective potential of
\eqn\effpot{V(R) \propto K - R}
where $R$ is the transverse separation of the branes and $K$ is a constant
of order one in string units.  The constant contribution to the potential
can be thought of as due to massive open string modes between the branes.
This shows that there is a repulsive force between 0- and 6-branes at 
short distance, which should match smoothly to the repulsive $1/R^2$ force
found at large distances in \lifsh, reflecting the fact that this system 
has no supersymmetry.

It will be important to know how many and what kinds of degenerate modes
of the string contribute to this effective result.  To determine this, 
we first consider the branes coincident with no velocity and analyze the
massless 0-6 string modes.  The Neveu-Schwarz sector has a positive zero
point energy \tasi\ and so contributes no massless modes.  The Ramond sector
bosons are massive, so the massless modes to consider are R fermions.  These
arise from fields with same-type boundary conditions, either
Neumann on both sides or Dirichilet on both sides.  There are four such
directions: the time direction and the three spatial directions transverse
to the branes.  Label these directions by
the index $\mu = 0,1,2,3$.  The rotational symmetry in these directions is
preserved, and so the zero modes of these R sector fermions are the 
same as for the open superstring in four dimensions; there are four states
which can be represented as a four-component Dirac spinor.  The
GSO projection removes half the states, leaving one
two-component Weyl spinor.

This mode should have an effective description
in terms of gauge theories on the D-branes.  It reduces to fermionic
quantum mechanics
on the 0-brane world line with the transverse rotational symmetry as an
internal gauge symmetry.  If the D-branes have a small separation, 
$Y^\mu$,
then it is reasonable to assume that the lowest mass state of the string
stretched between the branes is a minimally stretched string mode which
approaches zero energy as the separation goes to zero.  Thus, this mode 
can be described in terms of the effective gauge theory by adding to the the
effective action for the Weyl fermion a mass term proportional to $Y^\mu$;
let 
\eqn\effact{S_{eff} = \int dt \( i \bar{\psi} \pp_{t} \psi - \bar{\psi}Y^{\mu} \bar\sigma_{\mu} \psi \)}
where $\psi$ is the Weyl spinor representing the zero mode and the 
$\sigma^{\mu}$ are the the Pauli matrices.  This interaction term is the only
linear mass term which can be constructed from $Y^{\mu}$.

The states of this effective theory can now be determined.  For simplicity,
choose coordinates so that the branes are separated in the $z$ direction.  
This system
then contains two two-level systems composed of the two components of $\psi$
and their complex conjugates,
so there are four states.  However, because there is no field strength
for the gauge potential, $Y_{0}$, the equation of motion for $Y_0$ just
requires that all states be charge neutral.  Note for the case of only
one 0-, 6-brane pair this implies that the net 6-brane gauge charge is also
zero.  This constraint eliminates two
states, leaving two physical states.  The ground state is tachyonic with
mass $-Y_z$, and describes the linear repulsive potential found in 
\effpot .  However, the excited state has mass $Y_z$ and so results in 
a linear attractive potential between the D-branes.

Consider now the thought experiment of slowly moving a 0-brane by a 
stationary 6-brane by the application of some external force.
We have seen that the interaction is effectively described by a two level
system with mass proportional to the brane separation. This
experiment is then analogous to the 0-0 brane scattering experiment considered
in \douglas .  If the 0-brane approaches the 6-brane from the $+z$
direction with some
small impact parameter, then after the branes cross, assuming that no other
modes effect the interaction, the separation coordinate, $Y^{z}$, changes
sign.  Thus the ground state mass is positive after the scattering event.  This
is interpreted as an excited state of ``out'' modes defined with the 0-brane
infinitely separated in the $-z$ direction.  Therefore the two levels cross 
when the branes have zero separation.

Thus, in analogy to the 0-0 brane scattering case, it would seem that
a fermionic string mode is usually created when 0- and 6-branes collide.
This fermionic mode is long lived, assuming that the 0-brane has small
enough velocity that the lifetime estimates made in \douglas\ apply.  The
excited mode then causes the two D-branes to attract and recollide.  
The D-branes will recollide many times until the excited
fermionic mode decays, allowing them to escape to infinite separation.  This
unstable state is the analog of the 0-0 brane metastable state of \douglas .
Of course, the 0- and 6-brane can not collide from infinite separation 
at small velocity initially
because of the repulsive force between them.  However, this unstable
state can be constructed by imposing
initial conditions in which the D-branes are closely separated with small
relative velocity and the fermionic mode is in the excited state. 

The interest in this configuration here is in bound states of 0- and 6-branes
which could correspond to the black hole above.  Since this black hole
is extremal, one would like to find a metastable state where the D-branes
have minimal velocities.  What is the lower limit of the 0-brane velocity,
$v$, for this state to exist?  It is consistent for the 0-brane to have zero 
velocity.  In 
the two 0-brane case, because of 
supersymmetry, the leading order effective potential between the branes
is proportional to $v^4$.  Thus some velocity was necessary in order to bind
them together with open strings.  Also, because there is enhanced
symmetry when the two 0-branes coincide, additional massless modes exist for
some range of close separations.  In order to ignore the effects of these 
modes, a lower limit on the velocity was required.  For the 0-6 brane case,
there is no enhanced symmetry of this type when the branes coincide, and so
such a lower limit on the velocity cannot be made.

The upper limit on the
velocity is $v^2 < R$, where $R << 1$ is the transverse separation in string
units, so that
velocity-dependent contributions to the effective potential can be ignored.
Therefore a bound state might be formed by having 0- and 6-branes 
coincident with zero relative velocity but enough fermionic zero modes 
excited so that the D-branes experience an attractive potential.  This is
the sort of D-brane picture which will be presented in Sec. 6.

\newsec{Definition of Entropy and Correspondence}

Before one can consider finding a statistical description of the entropy
of this black hole in terms of D-branes, one must define entropy for
a non-equilibrium state.  If a state is near an equilibrium state in some
sense, one may define the entropy of the non-equilibrium state to be 
approximately that of the equilibrium state it approaches in some limit.
Consider the black hole solution and its thermodynamic entropy.  It is
unstable, but for large charges its decay rate is exponentially suppressed.
Thus, in the limit of infinite charges it is stable and one could compute the
Bekenstein-Hawking entropy in the usual way:  the horizon area over four.
For charges large but not infinite, the black hole is approximately in an 
equilibrium state if one ``turns off'' Hawking radiation.  This is usually
accomplished by considering putting the black hole in a box with thermal
gasses of 0- and 6- branes at an appropriate temperature to counteract the
Hawking radiation, and considering the black hole as a sub-system.
Operationally this is the same as ignoring the Hawking radiation and claiming
that the entropy calculated in the usual way has meaning because the state
is close to equilibrium.  This is what I wish to do here because it will
allow a reasonable definition of entropy for the D-brane configuration in
terms of turning off a small quantum effect.

The correspondence principle, were it to apply in this case, would relate 
the masses and entropies of the black hole and D-brane
picture up to factors of order one at the correspondence point.  For fixed
charges, this is the value of the string coupling at which the curvature at 
the black hole event horizon becomes one.  In this case, for large charges 
$Q_0 \approx Q_6 = Q$ this value is $g = 1/Q$.  At this point the black
hole mass is $M \approx Q^2$ and the entropy is $S_{BH} \approx Q^2$ up to
factors of order one.

\newsec{D-Brane Picture: Microstate Counting}

In this section I will argue for an effective D-brane picture of the black 
hole, identifying the low energy degrees of freedom which lead to a 
statistical description of the entropy and show that the mass and entropy
calculated in this way indeed match the black hole results \bhmass\ and 
\bhentropy\ at the correspondence point.

Consider a system of $Q_0$ 0-branes and $Q_6$ 6-branes close together with
small relative velocities.  This is an unstable configuration.  Because the
energy of the 0-6 fermionic string modes is much less than the energy in
the D-branes, the number of low energy 0-6 fermionic string modes attached 
to a particular
D-brane will fluctuate, and unless at least half
are in their excited state, there is a net repulsive
effective force on the D-brane.  The D-brane will then either escape to 
infinite distance from the rest of the system if there are no excited 
open string modes
attatched to it, or be bound at a separation of order one if there are.
This is because the long range effective potential between 0- and 6-branes 
falls off as $1/R$, a weaker repulsive force than the attractive force the
string mode creates.  If it becomes bound by long open string modes, 
then eventually these will decay, releasing the D-brane to infinite 
separation.  Either way, the collection of D-branes will eventually disperse. 

However, for large numbers of D-branes, each D-brane is connected by many
different open strings to other D-branes.  Thus the likelihood that the
number of fermionic open string modes in the excited state attached to a 
particular D-brane will fluctuate
to a low enough value for a long enough time that the D-brane can escape
the short separation regime is very small.  Therefore the state with many 
D-branes is metastable.

This metastable state is ``close'' to an equilibrium configuration in which
the small quantum effect of open string mode decay is neglected.  If such
an equilibrium configuration can be consistently constructed, then the 
logarithm of its ground state degeneracy will closely approximate the
statistical entropy of the D-brane picture corresponding to the black hole
under investigation.

How can one consistently describe such an equilibrium state?  Since open
string mode decay is neglected, the D-branes do not need relative velocities to
create more modes.  It is sufficient to have all the D-branes coincident
at zero velocity and choose the configuration of the fermionic open string
modes.  This is clearly a minimal energy configuration, since there is
no D-brane kinetic energy and all the open string modes are degenerate
and massless.
To make this a stable configuration, at least half the modes on each
D-brane must be in an excited state; then if one considers small 
perturbations of the positions of the branes, the effective potential is
always attractive.  The remaining modes can be chosen arbitrarily.

The physical states of this system are those configurations of modes
in the effective gauge theory which have zero net 0-brane gauge charge 
on each 0-brane and zero
net 6-brane gauge charge on each 6-brane.  The 6-brane gauge charge
must be conserved because the 6-branes are wrapped around compact 
directions so no flux can escape to infinity.  It was shown
above that this constraint can be met for each individual 0-brane,
6-brane pair leaving two physical states per pair.

There are actually more physical states than this because it is
possible to satisfy the charge neutral constraint on a particular
D-brane using modes from different D-branes.  However, including
these extra states does not change the leading order behavior of
the number of degenerate configurations. Because
the statistical entropy must match the thermodynamic entropy
only up to factors of order one, the leading order behavior of the
number of degenerate configurations is all that is required. Thus
it suffices to restrict attention to the $2 Q_0 Q_6$ states found
above.

Note that only modes from 0-6 strings have been considered; the open strings
between like D-branes have been ignored.  It seems likely that the presence
of both 0- and 6-branes will break the enhanced symmetry usually found
when many like D-branes are superimposed.  In addition, the usual modes will
be velocity suppressed in the physical non-equilibrium picture.  In any case,
these modes can not change the leading order behavior of the mass or entropy,
so it is consistent to omit them from consideration.

Now the number of consistent degenerate configurations can be
determined to leading order.  The contribution to the number
of configurations due to ways of choosing half of the two-level
systems associated with each D-brane to be in the excited state 
is sub-leading.  Therefore,
the leading order behavior is determined by choosing the level
of the remaining half of the two-level systems arbitrarily.
For large charges, $Q_0 \approx Q_6 = Q$,
the number of microstates to leading order at correspondence is
\eqn\nos{N \approx 2^{Q^2}}
where the numerical factor in the exponent has been dropped since it will
contribute a factor of order one to the entropy.
This gives an approximate statistical entropy at correspondence of
\eqn\statent{S_{stat} \approx \log{N} \approx Q^2 ,}
which matches the black hole entropy $S_{BH}$ up to factors of order one.

The total mass of this system is composed of three parts:  the masses of the
constituent D-branes, the 0-point energy due to massive 0-6 string modes, and
the mass of the light open string fermionic modes.
The following computations are valid at the correspondence point.  
The total mass of
the constituent D-branes is $\approx Q/g \rightarrow Q^2$.
The 0-point energy is order one in string units, and there are $\approx Q^2$ 
open strings, so the total 0-point energy contribution to the mass is 
$\approx Q^2$.  As the string coupling is increased to the correspondence
point, the D-branes ``spread out'' quantum-mechanically to about the
string scale. Thus, even though the low energy fermionic open string
modes are massless as $g \rightarrow 0$, they may contribute some mass
at the correspondence point.  The stretched open strings must remain 
shorter than order one, so their mass is at most of order one.  Note that 
this fact 
prevents the open strings from being longer than the 
Schwarzschild radius at correspondence.  
There are $\approx Q^2$ of these strings, so their 
contribution to the total mass is again $\approx Q^2$.  Thus the total
mass of this configuration is $\approx Q^2$ which agrees with the black hole
result \bhmass\ at corresopondence.

\newsec{Discussion}

An extremal, non-BPS black hole solution to the low energy effective
action of type IIA string theory with 0- and 6-brane charge has been presented.
It is interesting that a non-supersymmetric bound state of 0- and 6-branes
exists since the effective force between them is, at first glance,
repulsive at all scales.
The black hole is unstable to decay of 0- and 6-branes, but for large 
charges it is a
long-lived state.  The description of this state in terms of D-branes is
that $Q_0$ slowly moving 0-branes and $Q_6$ slowly moving 6-branes are bound 
by excited but low
energy fermionic modes
of strings connecting different branes.  The zero point energy of these
connecting strings, due to massive modes, contributes enough mass to the
configuration to account for the black hole's excess mass over the mass of the
constituents at correspondence.  The number of configurations of 
the approximately
massless fermionic modes gives the microstates from which the statistical
entropy of the system can be computed, and this entropy matches the
thermodynamic entropy of the black hole at correspondence.  This system is
then more evidence that the correspondence principle of \garyandjoe\ is
valid for black holes with two different RR charges.

There are two obvious generalizations of this solution.  The first is
just the non-extremal version with constant dilaton and the second is the
class which carries dilaton charge. 
For solutions far from extremality, the charges become unimportant and
the analysis for Schwarzschild in \garyandjoe\ applies.  For solutions
near extremality, 
one
problem is that
non-extremal black hole solutions can radiate closed strings as well as
D-branes and it is unclear how to generalize this D-brane picture to
include such processes.  The solutions which carry dilaton charge can be
determined by modifying the 5-D Kaluza-Klein black hole solutions with 
non-constant dilaton found in \gibwilt .

\vskip 1cm
\centerline{\bf Acknowledgements}
I would like to thank G. Horowitz, J. Polchinski, M. Douglas, J. Pierre, 
M. Srednicki, and P. Pouliot for discussions.
It was supported in part by NSF Grant PHY95-07065.
\listrefs

\end